\documentclass[prb,preprint]{revtex4-1} 

\usepackage{amsmath}  
\usepackage{amsfonts} 
\usepackage{graphicx} 

\begin{document}
\def\be{\begin{equation}}
\def\ee{\end{equation}}

\begin{center}
{\large\bf{Estimation of the error matrix in a linear least square fit to the data from an experiment performed by smartphone photography}}
\end{center}

\begin{center}
{Sanjoy Kumar Pal$^1$, Soumen Sarkar$^2$, Surajit Chakrabarti$^3$ \\
$^1$Anandapur H.S. School, Anandapur,Paschim Medinipur,WB,India\\
$^2$Karui P.C.High School, Hooghly, WB, India\\
$^3$Ramakrishna Mission Vidyamandira, Belur Math, Howrah, WB,India\\}
{E-mail:surnamchakrabarti@gmail.com}
\end{center}

\begin{center}
{\large\bf{Abstract}}
\end{center}

Determination of the Young's modulus of a metal bar in the form of a cantilever is an old experimental concept. However, we have taken the advantage of modern advanced technology of smartphone camera to find the  load depression graph of the cantilever by taking photographs with the smartphone camera. Smartphone photography allows us to find a precise transverse magnification of an object from the size of the real image formed on the sensor of the camera. Image size on the sensor can be obtained with micron level accuracy. From the load depression graph, we have determined the Young's modulus of the bar. The sensitive measurements of the depression of the cantilever at its free end by its own weight, have allowed us to determine the density of aluminium.
We have added an analysis of the chi squred minimisation technique for determining the parameters and their uncertainities in a linear fit. Starting from the curvature matrix we have made a comprehensive analysis of the error matrix relevant for a two parameter linear fit. Then we have shown how to form the error matrix for the fitted parameters which includes the covariance term between the two correlated parameters, in the context of our specific experiment. We have propagated the errors in the parameters  to find the uncertainties in the Young's modulus and the density of the bar.  We have shown that a precise measurement is possible by smartphone photography.

\newpage
\section{Introduction}

Young's Modulus of an aluminium bar in the form of a cantilever has been deternined in this work. We have also determined the density of aluminium. We have applied a modern technique for determining the depression of the free end of the cantilever as a function of load. In this method we have taken the photograph of the bent bar along with a horizontal laser beam which plays the role of a reference level. The photograph has been taken by a smartphone and analysing the photograph we have determined the depression of the free end of the bar from the reference level of the laser beam as a function of the load.

A smartphone apart from its communication pupose is now being used extensively for physics education as is evidenced by the number of journal publications from all over the world.$^1$ In a recent work$^2$ we have shown how to find the object distance and its transverse width from the smartphone photographs from two positions separated by a distance along the line of sight of the camera. In this work we have assumed that the thin lens equation works for a compound lens of the smartphone camera. In a recent work$^3$ questions have been raised whether a thin lens equation will work for the compound lens of the camera. We have clarified the issue by introducing the concept of equivalent lens.$^4$ We have shown that when a thin lens having the same focal length as that of the compound lens is placed at the first principal plane of the compound lens, then its magnification is same as that produced by the camera lens. This establishes that the thin lens analysis is perfectly valid for the compound lens of the smartphone camera. We have further validated our ideas by applying it to the virtual images produced by thin lenses.$^5$

In a recent work$^6$ Young's modulus of a steel ruler in the setting of a cantilever has been determined. The depression of the free end has been determined by noting the bent position of the cantilever against a calibrated linear potentiometer.

It is our experience that the undergraduate students generally do not have a clear idea of how to find the parameters and their uncertainties of a simple linear fit to a set of data. There are freely available software in the internet which do these fitting tasks for the students. However, students do not learn the concept of error matrix for the uncertainties. In sections III to VI, we have introduced the very basics of extracting the linear fit parameters by minimising the chi square and have shown how to get the error matrix including the off diagonal covariance term. In these sections we  have presented a short and comprehensive analysis of errors in the context of an undergraduate level experiment done in a novel manner using smartphone photography. We believe that a thorough understanding of the concept of error propagation is essential for the students to be able to design a new experiment so that the uncertainties in the results are minimal with the available instruments.

\noindent
\section{Theory for determining the Young's modulus and density of the aluminium bar}

In this work we have worked with an aluminium bar of rectangular cross section taken in the form of a cantilever of length $L$, breadth $b$ and depth $d$. When a load $Mg$ is suspended at the free end of the cantilever, the small depression $l$ at the free end can be related to the Young's modulus as$^6$ \be l=\frac{4gL^3}{bd^3Y}(M+\frac{3}{8} m_0).\ee
Mass of the cantilever part of the bar is $m_0.$ Using this equation we find the Young's modulus of the bar.

For the measurement of all the depressions, we take a horizontal beam of a laser as our reference line as shown in Fig.1. All depressions of the bar ($h_i$'s) at its free end are measured by photography method (to be described later) with respect to this laser beam. We determine the height of the laser beam from the horizontal part of the bar kept fixed on the table with two C-clamps. We call this height as $s$. So the actual depression $l_i=h_i-s.$

 Eq.(1) can be written as \be l=a_1+a_2M\ee where \be a_1=\frac{3gL^3}{2bd^3Y}m_0\ee and \be a_2=\frac{4gL^3}{bd^3Y}\ee with \be \frac{a_1}{a_2}=\frac{3}{8}m_0.\ee

We make a  least square fit to Eq.(2) where we plot $l_i$ as a function of $M_i$. We determine $a_1$ and $a_2$ from the fit and determine Young's modulus of the aluminium bar using Eq.(4) from the slope.

We find the mass per unit length of the bar $m_u$  from an equation which is obtained by looking at the small depression of the bar because of its own weight, when the suspended mass $M=0.$
From Eq.(1) we get the depression of the bar for different lengths of the cantilever due to its own weight as a function of length as \be l=\frac{3}{2}\frac{gm_u}{bd^3Y}L^4.\ee 
We get the depression of the bar without any added load for different lengths of the cantilever due to its own weight as a function of length  as \be l_i=a_2 L_i^4.\ee We plot $l_i$ as a function of $L_i^4$ and fit a straight line to the data. We determine $m_u,$ from the slope $a_2$,when $Y$ is  already known. Hence we find the density of the aluminium bar.


In section V  we give an analysis of how to determine the uncertainties of the fitted values of the parameters. In section VI we do the same thing when the straight line passes through the origin. We will show there that the determination of the density of the bar entails less uncertainty when it is determined from Eq.(7) rather than from Eq.(5).


\section {Chi-square analysis for a two parameter linear fit}

In this section we  discuss  the method of fitting  a two-parameter linear function 
\begin{equation}\label {fiteqn}
 y=a_1+a_2\,x\end{equation} 
to a set of  experimental data  by the method of minimum $\chi^2$.

Given a set of data $\{x_i,y_i\}$ with  errors in $y_i$ 
denoted by $\sigma_i$, we define a quantity called 
$\chi^2$ as
\begin{equation} \label{chisquare}
\chi^2\equiv
\sum_i \frac{\left[y_i-(a_1+a_2\,x_i)\right]^2}{\sigma_i^2}\,.
\end{equation}This definition of $\chi^2$ is valid only when the variables $y_i$ are uncorrelated. 
We can determine the parameters $a_1$ and $a_2$ by minimizing the  $\chi^2$ with respect to $a_1$ and $a_2$. 
By setting $\partial \chi^2/\partial a_1$ and 
$\partial \chi^2/\partial a_2$ to zero, we find
\begin{align} 
a_1 &= \frac{GC-HB}{AC-B^2}\,,
\\
a_2 &= \frac{HA-GB}{AC-B^2}\,,
\end{align} 
where 
\begin{align} 
A&=\Sigma_i \frac{1}{\sigma_i^2}\,,
\\
B&=\Sigma_i \frac{x_i}{\sigma_i^2}\,,
\\
C&=\Sigma_i \frac{x_i^2}{\sigma_i^2}\,,
\\
G&=\Sigma_i \frac{y_i}{\sigma_i^2}\,,
\\
H&=\Sigma_i \frac{x_iy_i}{\sigma_i^2}\,.
\end{align}
$a_1$ and $a_2$ can be determined using  Eqs.(10) and (11).

\section {Determination of the variances of the parameters and the covariance between them}

 Our main purpose in this section is to determine the variances and the covariance between $m$ parameters.$^{8-11}$ The $\chi^2$ as introduced in the last section is now written as, 

\be \chi ^2=\Sigma _{i=1}^n\frac{1}{\sigma_i^2}[y_i-f(x_i,\vec a)]^2\ee where [$a_k$] are found by solving the equations \be\frac {\partial \chi^2}{\partial a_k}=0,\ee where $1\le k\le m.$ The variance of the parameters is given by$^{11}$
\be \sigma^2_{a_k}=\Sigma_{i=1}^n \sigma^2_i (\frac{\partial a_k}{\partial y_i})^2\ee and the covariance term$^{12}$\be \sigma_{a_ka_l}=\Sigma_{i=1}^n \sigma^2_i (\frac{\partial a_k}{\partial y_i})(\frac{\partial a_l}{\partial y_i}).\ee We assume all the measurements $y_i$ are independent and $\sigma_i$'s are small. Eq.(18) can be considered to be an implicit relationship between $a_j$'s and $y_i$'s. We take a second derivative of $\chi^2$ in Eq.(18) with respect to $y_i$ and this should also be zero. Remembering $\chi^2$ has an implicit dependence on $y_i$ via $a_k$'s, we get,
\be \frac {\partial^2\chi^2}{\partial y_i\partial a_k}+\Sigma_{j=1}^m \frac {\partial^2\chi^2}{\partial a_k\partial a_j}(\frac {\partial a_j}{\partial y_i})=0.\ee We define the curvature matrix $\alpha$ by \be \alpha_{kj}=\frac{1}{2}\frac {\partial^2\chi^2}{\partial a_k\partial a_j}.\ee Eq. (21) can be written as a matrix equation \be \Sigma_{j=1}^m\alpha_{kj}\frac{\partial a_j}{\partial y_i}=-\frac{1}{2}\frac {\partial^2\chi^2}{\partial y_i\partial a_k}\ee with $\frac {\partial  a_j}{\partial y_i}$ forming a column vector with different $a_j$'s. We can solve Eq.(23) and write \be \frac {\partial  a_j}{\partial y_i}=-\frac{1}{2}\Sigma _{l=1}^m\epsilon_{jl}\frac {\partial^2\chi^2}{\partial y_i\partial a_l}\ee where $\epsilon$ is the inverse of $\alpha$ matrix. From the expression of $\chi^2$ in Eq.(17) we get \be \frac {\partial^2\chi^2}{\partial y_i\partial a_l} =-\frac{2}{\sigma_i^2}\frac{\partial f(x_i,\vec a)}{\partial a_l}.\ee

So we finally get the covariance term \be  \sigma_{a_ja_l}=\Sigma_{i=1}^n \sigma^2_i (\frac{\partial a_j}{\partial y_i})(\frac{\partial a_l}{\partial y_i})=\Sigma_{i=1}^n\frac{1}{\sigma_i^2}(\Sigma_{k=1}^m\epsilon_{jk}\frac{\partial f(x_i,\vec a)}{\partial a_k})(\Sigma_{p=1}^m\epsilon_{lp}\frac{\partial f(x_i,\vec a)}{\partial a_p}).\ee  

One can easily show  from the definition of the $\alpha$ matrix Eq.(22), $\alpha_{kp}= \Sigma_{i=1}^n \frac{1}{\sigma_i^2}\frac{\partial f(x_i,\vec a)}{\partial a_k}\frac{\partial f(x_i,\vec a)}{\partial a_p}.$ This equation is true only when $f(x_i,\vec a)$ is a linear function of $a_j$'s.

We are led to the following equation,\be \sigma_{a_ja_l}=\Sigma_{k,p=1}^m \epsilon_{jk}\epsilon_{lp}\alpha_{kp}=\epsilon_{jl}.\ee Since $\epsilon$ and $\alpha$ are inverse to each other, the sum $\Sigma_{k=1}^m\epsilon_{jk}\alpha_{kp}=\delta_{jp}$ where $\delta_{jp}$ is the Kronecker delta.

When $j$ and $l$ are same, we get \be \sigma_{a_j}^2=\epsilon_{jj}.\ee So, the diagonal terms of the error matrix $\epsilon$ give the variance of the parameters $a_j$ and the off-diagonal terms give the covariance between the two different parameters  as in Eq.(27).

\section {Error matrix and uncertainties in functions of the fitted parameters in a two parameter fit}

We can find the uncertainties in the parameters $a_1$ and $a_2$ following the
compact notation of Burrel.$^{11}$  We first form the curvature 
matrix $\alpha$, which is a $2\times 2$ matrix since we 
have only two parameters $a_1$ and $a_2$. 
The matrix elements are defined by 
\begin{equation}
\alpha_{ij}=\frac{1}{2}\frac{\partial^2 \chi ^2}{\partial a_i\partial a_j}\,,
\end{equation} 
where $i,j$ each 1can take only two possible values 1, 2.  
It can easily be shown that 
\begin{equation}\label{alpha}
\alpha=
\begin{pmatrix}
	 A& B\\
	B& C \\
\end{pmatrix}\,,
\end{equation}
where $A$, $B$, and $C$ are defined in Eqs.(12)-(14). 

We have shown in Eq.(24),
\begin{equation} 
\epsilon= \alpha ^{-1}\,.
\end{equation} 
It was also shown in Eq.(28) that for a linear fit, the uncertainties in the estimates of $a_1$ and $a_2$ 
are given by $\sqrt {\epsilon_{11}}$ and  $\sqrt {\epsilon_{22}}$,
respectively.
However, the parameters  $a_1$ and $a_2$  would be correlated as both are being determined 
from  the same set of data. Off-diagonal terms 
of the  matrix $\epsilon$  would give this correlation. It is to be noted that the parameters $a_1$ and $a_2$
would be correlated even when the $y_i$ variables are uncorrelated. We can easily determine the matrix $\epsilon$
 by using the standard formula 
for the inverse of a matrix as given by
\begin{equation} 
\alpha^{-1}=\frac {\text {adj }\alpha}{\text {det } \alpha}\,,
\end{equation} 
where $\text {adj}$ refers to the adjoint matrix and $\det$ 
refers to the determinant of the matrix $\alpha$. We finally get
\begin{equation}\label{epsilon}
\epsilon=\frac{1}{\text{det } \alpha}
\begin{pmatrix}
	 C& -B\\
	-B& A \\
\end{pmatrix}\,,
\end{equation}

In order to determine the uncertainty of any function $g$ that
depends on $a_1$ and $a_2$,
we have to propagate the errors as discussed by Tellinghuisen.$^{13}$ We form the column vector, 
\begin{equation}
V=\begin{bmatrix}
{\partial g}/{\partial a_1}  
\\
{\partial g}/{\partial a_2}
\\
\end{bmatrix}\,.
\end{equation}
The variance term for $g$  can then be 
written in  compact notation as,
\begin{equation}\sigma_g^2= V^T\epsilon V\,,
\end{equation}
where $V^T$ is the transpose of the  column matrix $V$ and we get,

\begin{equation}\label{sigfsq}
\sigma_g^2= \left(\frac{\partial g}{\partial a_1}\right)^2\sigma_{a_1}^2 + 
\left(\frac{\partial g}{\partial a_2}\right)^2\sigma_{a_2}^2
+2\sigma_{a_1a_2}\,\frac{\partial g}{\partial a_1}\frac{\partial
g}{\partial a_2}\,.
\end{equation}
If $a_1$ and $a_2$ are uncorrelated, the covariance term drops out.  
\section { Estimation of the fitted parameter and its uncertainty for a plot passing through the origin}

We get the depression of the bar for different lengths of the cantilever due to its own weight as a function of length of the cantilever as shown in Eq.(7).
We determine $a_2$ and hence mass per unit length of the cantilever $m_u$ using Eq.(6), when $Y$ is already determined.

 \be \chi^2=\Sigma_i^n\frac{1}{\sigma_i^2}(y_i-a_2x_i)^2\ee where $x_i=L_i^4,$ and $y_i=l_i$.
 We minimise the $\chi^2$ with respect to $a_2$ and get the equation
\be a_2= \frac{\Sigma_i \frac{x_iy_i}{\sigma_i^2}}{\Sigma_i \frac{x_i^2}{\sigma_i^2}}.\ee
\be \delta a_2=\frac{1}{\Sigma_i \frac{x_i^2}{\sigma_i^2}}
(\Sigma_i (\frac{x_i}{\sigma_i^2})^2\delta y_i^2)^{\frac{1}{2}}
=\sqrt{\frac{1}{\Sigma_i \frac{x_i^2}{\sigma_i^2}}}.\ee

\noindent
\section{Principle of determining the depression of the bar by photography}

Our aim in this section is to find the depression of the free end of the cantilever bar under a load as measured from a horizontal laser beam set as a reference where the beam is vertically above the bar. We take two photographs from two positions of the camera separated by a distance $D$ along the line of sight of the camera. Each photograph shows the bent bar and the laser beam. We describe below how the depression can be determined by analysing the photographs. This has been described in greater detail in a recent work.$^2$ 


When an image is formed by a thin lens, we have the relation$^2$
\be\frac {1}{v}-\frac {1}{u}=\frac {1}{f_c}.\ee Here,
$u$ is the object distance and $v$ is the image distance measured from the lens. The focal length of the lens $f_c$ is positive for a convex lens according to our sign convention. When a real image is formed by a convex lens, object and image positions are on the opposite sides of the lens and the image is inverted. Transverse magnification produced by the lens is given by \be m=\frac{v}{u}=\frac{I}{O}\ee where $I$ is the image size on the sensor of the camera and $O$ denotes the distance between the laser beam and the red mark at the lower end of the bar.
From the above two equations we have \be m=\frac{1}{1+\frac{u}{f_c}}.\ee Inverting this equation we get \be f_c=\frac{u}{\frac{1}{m}-1}.\ee  For real image, $u$ and $m$ are negative according to our sign convention. So, \be f_c=\frac{\left | u\right |}{\frac{1}{\left | m\right |}+1}.\ee 

It has been shown in a previous work,$^4$ a thin lens is equivalent to a compound lens when the thin lens is placed at the first principal plane of the compound lens and its focal length is same as that of the compound lens. In this condition the transverse magnification of the compound lens is same as the magnification of the thin lens for object at the same position. For this reason we can use the thin lens equation for the analysis of the camera image.

If $u_1$ and $u_2$ are the  distances of the cantilever bar from two positions of the camera with  transverse magnifications $m_1$ and $m_2$ respectively, we get \be D=f_c(\frac{1}{m_2}-\frac{1}{m_1})\ee where $D=u_2-u_1$.
Hence we get, \be O=\frac{D}{f_c(\frac{1}{I_2}-\frac{1}{I_1})}.\ee  For simplicity  in calculation we can write \be O=\frac{\left |D\right |}{f_c\left |(\frac{1}{I_2}-\frac{1}{I_1}\right)|}.\ee 


\section{Experiment and the Results}
\subsection{Experimental Setup}
We have measured the Young's Modulus of an Aluminium bar taken as a cantilever as shown in Fig.1. The horizontal laser beam serves as our reference. The load is suspended at the free end of the bar. The other end of the bar is placed on a horizontal table  clamped by two C-clamps. A photograph of the bar and the laser beam is shared with the Apple Operating System(OS). The photograph  can be opened by the software 'Preview'. By setting the cursors at the two points between which we need the distance, the pixel count along the X and Y axes are given by the software.
The pixel count is converted into mm from which we get the transverse magnification of the real image on the photo sensor of the camera.The dimension of 1 pixel is available from the relevant website.$^{14}$ The focal length of the camera can also be found from the relevant website.$^{15}$ The detals of the method for determining the magnification from the photograph has been discussed elsewhere.$^{2,4}$
\subsection{Camera specifications}

 We have used a smartphone  model iPhone 12 Pro Max. In the camera we have used the standard wide lens option with ``$\times 1$" magnification. The camera  least count $R=1.7\mu$m, which is the length of one pixel on the photosensor of the smartphone. The focal length of the camera lens is $0.532\pm 0.0004$cm as we found out in our work.$^4$ We use the focal length found out by us rather than the focal length given in the website.$^{15}$
\subsection{Determination of the depth($d$) of the experimental bar by photography, and the breadth by a digital slide calliper}

 We use the photographic method to determine the depth of the aluminium bar  using Eq.(44). We have taken a substrate and pasted a graph sheet on it. We take its photograph keeping the smartphone vertically above the graph sheet, held by a stand. By analysing the photograph by the software 'Preview' we determine the number of pixels that correspond to a length 10cm on the graph sheet. From the pixel data we calculate the image width $I_1$ on the sensor of the camera. Then we remove the graph sheet and place the bar on the substrate. We then paste the graph sheet on the bar and take a photograph from the same place of the smartphone. We call the image size on the sensor $I_2$. 
We show the data in Table I. We determine d using the relation \be d=u_1-u_2=f_c\times O(\frac{1}{I_1}-\frac{1}{I_2})\ee where $O$ is the size of the object which in our case was 10.0cm and $I_1$ and $I_2$ are the image sizes on the sensor of the camera, where $I_1=\text {pixel1}\times R$ and similarly for $I_2$. Typically we have a maximum uncertainity of 2 pixels in repeated measurement of a pixel count. We find the average thickness of the bar as $d=0.217\pm 0.001$cm

The breadth of the bar was found by a digital slide calliper. It was found to be 2.39cm.
\subsection{Determination of height of the reference laser beam from the horizontal part of the beam resting on a table}

We have first levelled the working table carefully and found the height of the laser beam from the mark at the bottom of the part of the aluminium bar which is resting on the table. 

After we take a photograph of the aluminium bar and the laser beam with a smartphone, we can share it with a computer and can analyze the photograph  with the `Preview' software with Apple Operating System(OS) (Software `Paint' in the Windows OS). When the cursors of the Preview software are coincided with the two edges of the object $O$, it gives the pixel count for the width of the real image on the sensor of the camera. We shift the camera by a distance $D$ and find the image widths $I_1$ and $I_2$ on the sensor of the camera. We determine the width of the object $O$ using Eq.(47) if the focal length of the camera lens is known before hand.   
 The data are shown in Table II. The average of these heights turns out $s=2.156\pm 0.001$cm
 
\subsection{Determination of the Young's Modulus Y} 
We determine the Young's modulus of the aluminium bar by photographing the depressed bar with load along with the laser reference line as seen in Fig.1. We use 6 loads and go on increasing the load and take photographs for each load. Then we do the same thing while decreasing the load. Then we shift back the camera by a suitable distance D along the line of sight of the camera. We use the same set of loads and take another set of  photographs. We then analyse the photographs off-line by sharing the photos with the 'Preview' software of the Apple OS(the software is 'Paint' for Windows OS). We find the pixel difference on the sensor between the laser line and the red mark on the bar at the point where the load is suspended. We determine the  depressions $h_i$ as described in section VII using Eq.(47). In Tables III and IV we show the detail data for two lengths of the calntilever part of the beam. We have taken 6 different lengths of the cantilever. Actual depression of the bar is $l_i=h_i-s$ where $s$ and its uncertainty have  been presented in the subsection D.

 Uncertainties in these values of $l_i$ are found by propagating the uncertainties in each term of Eq.(47) as follows:
\be \frac{\delta h_i}{h_i}= \sqrt{(\frac{\delta D}{D})^2+\frac{1}{(1-\frac{I_2}{I_1})^2}(\frac{\delta I_2}
{I_2})^2+\frac{1}{(\frac{I_1}{I_2}-1)^2}(\frac{\delta I_1}
{I_1})^2}.\ee Finally, we get \be \delta l_i=(\delta h_i^2+\delta s^2)^{\frac{1}{2}}.\ee In Figs.2 and 3 we show the load-depression plots where we plot $l_i$ along the Y axis along with their uncertainties for different loads plotted along the X-axis. We do this for all 6 lengths of the cantilever.

We have fitted Eq.(2) to the data and extracted $a_1$ and $a_2$ and the error matrix. We show the data and the value of Y in Table V. The average value of the Young's Modulus turns out to be $(7.03\pm 0.03)\times 10^{11}$ dynes/$\text{cm}^2$ where the uncertainty quoted is the standard error.
\subsection {Depression of the bar by its own weight with $M=0$}
 We photograph the bar without applying any load and bent by its own weight. This has been done for 6 different lengths of the bar. This data is shown in Table VI. In Fig.4 we show the plot of the depression $l_i$ as a function of $L^4$ according to Eq.7. We also show the fitted line along with the data points. Slope of the line turns out $a_2=(1.21\pm 0.02)\times 10^{-7}\text{cm}^{-3}$ From this we determine the density of aluminium to be $2.72\pm 0.06$ gm/cc.
\section{Conclusions}

In this work we have shown how to estimate the error matrix in a specific experiment in a very concise manner using the method of $\chi^2$ minimisation. We have included the covariance term between the  parameters in a two parameter linear least square fit. For a linear plot passing through the origin we  give the expression for the slope and its uncertainty from a one parameter least sqare fit.

 We believe that the detailed procedure when followed, will help students to design their experiment so that the results they find have minimum uncertainty. This endeavour will help them to design a new experiment optimally as one can directly see how the uncertainties are entering in different parts of the experiment. 

We show all these technical matters in the context of an experiment done at the undergraduate level, however with a new method for determining the depression of a bar as a result of the application of a load at the end of a cantilever. The depression has been measured by the method of smartphone photography. We have shown that the measurements are very precise mainly because of the fact that the modern technology allows us to measure the real images of an object on the sensor of the smartphone camera with a micron level accuracy.


\begin{table}[ht]
\centering
\caption{Data for the determination of the thickness(d) of the aluminium bar. Photographs taken from the same position of the camera.\\ Object size on the graph paper=10cm.}
\begin{tabular}{cccc}
\hline
\hline
obs&$I_1$&$I_2$&$d$\\
no.&cm&cm&cm\\
\hline
1&0.3850&~~0.3912&~~$0.216\pm 0.017$\\
2&0.4488&~~0.4571&~~$0.216\pm0.012$\\
3&0.4139&~~0.4211&~~$0.218\pm 0.015$\\
4&0.4687&~~0.4779&~~$0.218\pm 0.011$\\
5&0.4833&~~0.4930&~~$0.216\pm 0.011$\\
6&0.3725&~~0.3782&~~$0.218\pm 0.018$\\

\hline
\end{tabular}
\end{table}


\begin{table}[ht]
\centering
\caption{Data for the determination of the height of the laser beam above horizontal part of the aluminium bar. Photographs taken from two positions of the camera with displacement $D$.}
\begin{tabular}{ccccc}
\hline
\hline
obs& $I_1$ &$I_2$ &$D$&$s$\\
no.&(cm)&(cm)&(cm)&(cm)\\
\hline
1&0.1120&~~0.0287&~~29.7&~~$2.154\pm 0.036$\\
2&0.1010&~~0.0264&~~32.2&~~$2.163\pm 0.039$\\
3&0.1051&~~0.0277&~~30.5&~~$2.156\pm 0.037$\\
4&0.1076&~~0.0270&~~31.8&~~$2.154\pm 0.038$\\
5&0.1005&~~0.0277&~~30.0&~~$2.156\pm 0.038$\\
6&0.1056&~~0.0275&~~30.8&~~$2.153\pm 0.037$\\

\hline
\end{tabular}
\end{table}

\begin{table}[ht]
\centering
\caption{Data for the determination of the depression of the aluminium bar as a function of the load. Image sizes  $I_1$ and $I_2$ on the camera sensor are estimated from pixel data of photographs taken from two distances separated by $D=20.6$cm.\\ L=30.8cm; s=2.156cm}
\begin{tabular}{c|ccc|ccc|cc}
\hline
\hline
&&load&&&Load &&&\\
&&increasing&&&decreasing&&&\\
\hline
load&$I_1$&$I_2$&$h_1$&$I_1$ &$I_2$&$h_2$&$\bar h$&$l=\bar h-s$\\
gm&cm&cm&cm&cm&cm&cm&cm&cm\\

\hline
24.81&0.1210&0.0413&2.428&0.1212&0.0413&2.426&~~2.427&~~$0.271\pm 0.027$\\
45.54&0.1282&0.0440&2.594&0.1282&0.0440&2.594&~~2.594&~~$0.438\pm 0.028$\\
66.32&0.1358&0.0466&2.747&0.1358&0.0468&2.765&~~2.756&~~$0.600\pm 0.029$\\
87.22&0.1433&0.0488&2.865&0.1431&0.0488&2.867&~~2.866&~~$0.710\pm 0.029$\\
107.92&0.1506&0.0513&3.013&0.1506&0.0513&3.013&~~3.013&~~$0.857\pm 0.030$\\
128.21&0.1586&0.0534&3.117&0.1586&0.0534&3.117&~~3.117&~~$0.961\pm 0.030$\\

\hline
\end{tabular}
\end{table}

\begin{table}[ht]
\centering
\caption{Data for the determination of the depression of the aluminium bar as a function of the load. Image sizes  $I_1$ and $I_2$ on the camera sensor are estimated from pixel data of photographs taken from two distances separated by $D=21.3$cm.\\ L=41.8cm; s=2.156cm}

\begin{tabular}{c|ccc|ccc|ccc}
\hline
\hline
&&Load&&&load&&&&\\
&&increasing&&&decreasing&&&&\\
\hline
load&$I_1$&$I_2$&$h_1$&$I_1$ &$I_2$&$h_2$&$\bar h$&$l=\bar h-s$\\
gm&cm&cm&cm&cm&cm&cm&cm&cm\\
\hline
24.81&0.1627&0.0512&2.991&0.1625&0.0512&2.993&~~2.992&~~$0.836\pm 0.028$\\
45.54&0.1814&0.0570&3.328&0.1812&0.0570&3.330&~~3.329&~~$1.173\pm 0.030$\\
66.32&0.2011&0.0627&3.648&0.2011&0.0627&3.648&~~3.648&~~$1.492\pm 0.032$\\
87.22&0.2190&0.0683&3.974&0.2188&0.0683&3.976&~~3.975&~~$1.819\pm 0.033$\\
107.92&0.2378&0.0748&4.369&0.2378&0.0748&4.369&~~4.369&~~$2.213\pm 0.035$\\
128.21&0.2557&0.0802&4.678&0.2557&0.0802&4.678&~~4.678&~~$2.522\pm 0.037$\\

\hline
\end{tabular}
\end{table}

\begin{table}[ht]
\centering
\caption{Determination of the Young's modulus Y}
\begin{tabular}{cccc}
\hline
\hline
$L$&$a_1$&$a_2$&$Y$\\
cm &cm&cm/gm&$10^{11}\text{dynes/cm}^2$\\
\hline
41.8&~~$0.414\pm 0.029$&~~$0.0164\pm 0.0004$&~~$7.14\pm 0.19$\\
40.0&~~$0.282\pm 0.029$&~~$0.0148\pm 0.0004$&~~$6.92\pm 0.19$\\
37.7&~~$0.247\pm 0.028$&~~$0.0123\pm 0.0003$&~~$6.97\pm 0.22$\\
35.6&~~$0.218\pm 0.034$&~~$0.0103\pm 0.0004$&~~$7.05\pm 0.30$\\
33.3&~~$0.190\pm 0.026$&~~$0.0084\pm 0.0003$&~~$7.10\pm 0.29$\\
30.8&~~$0.128\pm 0.027$&~~$0.0067\pm 0.0003$&~~$7.03\pm 0.36$\\

\hline
\end{tabular}
\end{table}

\begin{table}[ht]
\centering
\caption{ Data for the determination of the depression of the cantilever bar by its own weight. $s=2.156$ cm}
\begin{tabular}{cccccc}
\hline
\hline
$L$&D&Image1&Image2&h&$l=h-s$\\
cm&cm&cm&cm&cm&cm\\
\hline
42.0&~~21.1&~~0.1397&~~0.0439&~~2.539&~~$0.38\pm 0.03$\\
45.1&~~21.1&~~0.1352&~~0.0447&~~2.648&~~$0.49\pm 0.04$\\
48.0&~~20.9&~~0.1403&~~0.0473&~~2.803&~~$0.65\pm 0.04$\\
51.1&~~20.9&~~0.1493&~~0.0503&~~2.980&~~$0.82\pm 0.04$\\
54.0&~~20.8&~~0.1605&~~0.0541&~~3.191&~~$1.03\pm 0.04$\\
56.9&~~20.5&~~0.1726&~~0.0587&~~3.428&~~$1.27\pm 0.04$\\

\hline
\end{tabular}
\end{table}



\begin{figure}[h!]
\centering
\includegraphics[width=14cm]{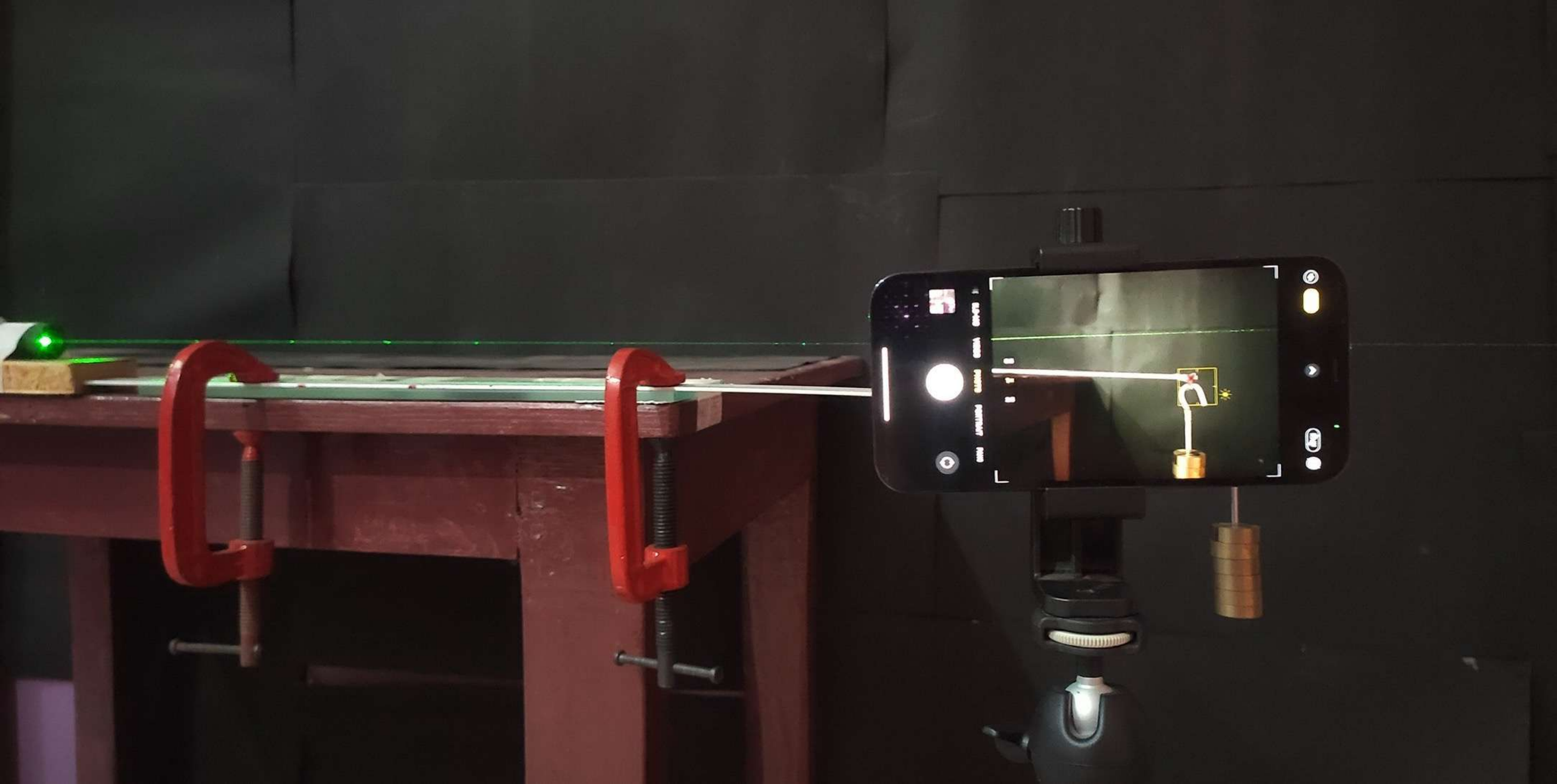}
\caption{Experimental setup}
\end{figure}

\begin{figure}[h!]
\centering
\includegraphics[width=13cm]{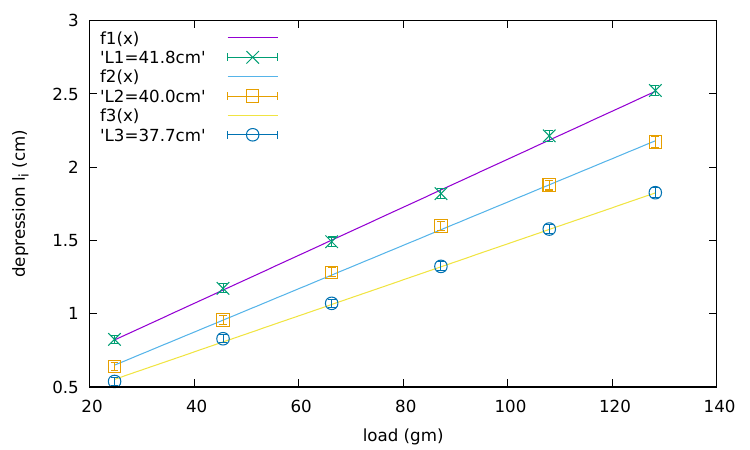}
\caption{Load-depression graphs for three different lengths of the cantilever }
\end{figure}

\begin{figure}[h!]
\centering
\includegraphics[width=13cm]{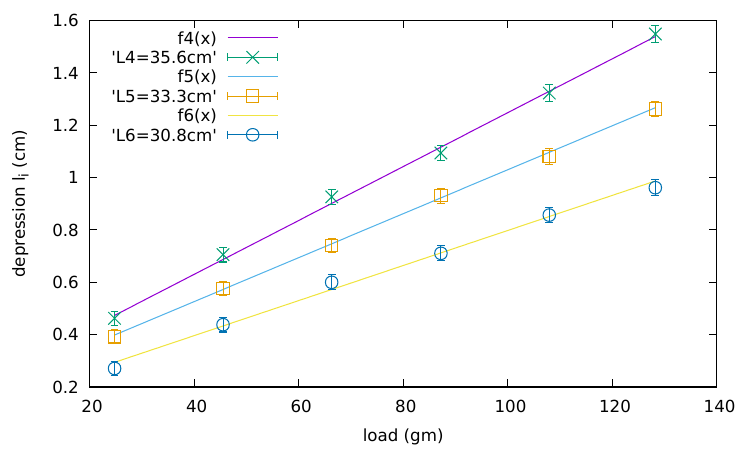}
\caption{Load depression graphs for three other lengths of the cantilever}
\end{figure}

\begin{figure}[h!]
\centering
\includegraphics[width=13cm]{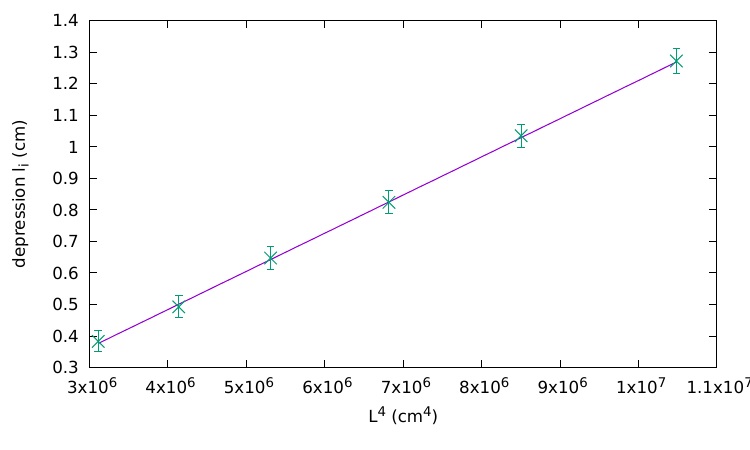}
\caption{Depression of the bar due to its own weight(External load zero) as a function of the length$^4$ of the cantilever}
\end{figure}

\end{document}